\documentclass[preprint2]{aastex6}

\usepackage{bm}
\usepackage{amsmath}
\usepackage{natbib}
\usepackage{multirow}
\usepackage{here}
\begin{document}


\title{DOUBLE-CELL TYPE SOLAR MERIDIONAL CIRCULATION BASED ON MEAN-FIELD HYDRODYNAMIC MODEL}


\author{Y.Bekki\altaffilmark{1} and T.Yokoyama\altaffilmark{1}}
\affil{Department of Earth and Planetary Science, The University of Tokyo, 7-3-1 Hongo, Bunkyo-ku, 113-0033, Japan}







\altaffiltext{1}{bekki@eps.s.u-tokyo.ac.jp}

\begin{abstract}
  
  The main object of the paper is to present the condition of the non-diffusive part of the Reynolds stress for driving the double-cell structure of the solar meridional circulation, which has been revealed by recent helioseismic observations.
By conducting a set of mean-field hydrodynamic simulations, we confirm for the first time that the double-cell meridional circulation can be achieved along with the solar-like differential rotation when the Reynolds stress transports the angular momentum upward in the lower part and downward in the upper part of the convection zone.
It is concluded that, in a stationary state, the accumulated angular momentum via the Reynolds stress in the middle layer is advected to both the upper and lower parts of the convection zone by each of the two meridional circulation cells, respectively.

\end{abstract}




\section{Introduction} \label{sec:intro}

The internal structures of the large-scale solar convections have recently been revealed in detail by helioseismic observations. Robust observational profile is obtained for differential rotation with a pole-equator difference of about $30\%$ of the rotation rate of the radiative zone \citep{thompson2003,howe2009,howe2011}. As for meridional circulation, however, the only robust observation is a poleward flow at the surface with the flow speed of $10 - 20$ $\mathrm{m \ s^{-1}}$ \citep{giles97,braunfan98,haber02,zhao04,gonzales08}. Although it is believed that the equatorward return flow exists inside the convection zone to meet the mass conservation, the exact depth of the return flow has been largely unknown.

Recently, \citet{zhao2013} reported detecting the equatorward flow in the middle depth of the convection zone, between $0.82 \ R_{\odot}$ and $0.91\ R_{\odot}$, and poleward flow again below $0.82 \ R_{\odot}$, where $R_{\odot}$ is the solar radius.
This observation suggests the double-cell structure of the solar meridional circulation with the counterclockwise circulation cell in the upper convection zone (poleward near the surface) and the clockwise circulation cell in the lower layer (poleward near the base). Moreover, \citet{kholikov2014} used other helioseismic measurements and also detected the evidence that the latitudinal flow changes its direction at several depths of the solar convection zone, supporting \citet{zhao2013}'s finding that the meridional flow consists of radially distinct circulation cells. 
However, poleward meridional flow near the base of the convection zone of this double-cell type circulation structure is problematic for the conventional flux-transport dynamo model \citep{hazra2014} because, in this model, the observed equatorward migration of sunspot groups is mainly attributed to the equatorward transport of toroidal magnetic fluxes by the meridional flow at the base of the convection zone where convective stable layer can detain magnetically buoyant toroidal fluxes from rising upward for long periods of time \citep{wang89,choudhuri95,dikpati99}.
Although the reliability of these local helioseismic measurements in the deeper convection zone is controversial and, in fact, there exists a contradictory observation which suggests the single-cell meridional circulation structure \citep{rajaguru2015}, theoretical investigation on the double-cell type meridional circulation should be regarded as of great importance as a first step to reconsider the validity of the conventional flux-transport dynamo model.

In this paper, therefore, we discuss the possibility of the double-cell meridional circulation in the framework of the mean-field theory. In a mean-field model, the effect of the angular momentum transport by the non-diffusive part of the Reynolds stress, commonly known as the turbulent angular momentum transport or the  $\Lambda$-effect \citep[see][]{rudiger93,kichatinovrudiger93}, is parameterized so that the differential rotation and meridional circulation are calculated self-consistently.
In fact, the parameterization of the $\Lambda$-effect contains some uncertainties.
In most of the previous mean-field simulations whose main interests focus on the Taylor-Proudman balance of the differential rotation \citep{kitchatinov1995,rempel2005}, the turbulent angular momentum fluxes are set nearly equatorward just in order to accelerate the equatorial region and to obtain the solar-like differential rotation. 
Although it is shown in these previous papers that the meridional circulation is very sensitive to the parameterization of the $\Lambda$-effect, the relation between the $\Lambda$-effect and the structure of the meridional circulation has not yet been investigated in detail.
Recently, \citet{pipin2016} studied the latitudinal dependence of the radial component of the turbulent angular momentum flux and produced the triple-cell meridional circulation in the mean-field framework.
However, the double-cell structure as suggested by \citet{zhao2013} was not reproduced and therefore the condition of the Reynolds stress for obtaining the double-cell meridional circulation is still unclear.

The relationship between the meridional flow profile and the azimuthal components of the Reynolds stress is described in terms of the so-called ``gyroscopic pumping'' \citep{miesch2011,featherstone2015}, which has been commonly used for explaining the formation mechanism of the solar near-surface shear layer \citep{hotta2015}.
In this paper, we first apply the equation of gyroscopic pumping for the double-cell meridional circulation and derive the condition of the Reynolds stress needed for obtaining the double-cell structure.
In general, however, it is not obvious whether the $\Lambda$-effect which reflects such a condition of the Reynolds stress can drive both the double-cell meridional circulation and the solar-like differential rotation at the same time, since the differential rotation and meridional circulation are hydrodynamically balanced with each other. 
We therefore conduct a set of mean-field hydrodynamic simulations to see whether or not the double-cell meridional circulation is hydrodynamically compatible with the solar-like differential rotation.

The organization of this paper is as follows.  In Section \ref{sec:two}, the condition of the Reynolds stress which is necessary for the double-cell meridional circulation is derived by considering the equation of gyroscopic pumping. In Section \ref{sec:three}, our mean-field model is explained.
The $\Lambda$-effect is specified in Section \ref{subsec:lambda-modeling}.
In Section \ref{sec:four}, we show the results of our mean-field simulations. Reference solution is presented in Section \ref{subsec:drmc} to confirm that the double-cell meridional circulation can be achieved along with the solar-like differential rotation.
The influence on the Taylor-Proudman balance is addressed in Section \ref{subsec:TPbalance}.
We further discuss the angular momentum balance in Section \ref{sec:refgyro} where the maintenance mechanism of the double-cell meridional circulation is explained.
In Section \ref{sec:validity}, we discuss the validity of our parameterization of the $\Lambda$-effect.
Finally, we conclude in Section \ref{sec:summary}.

\section{Gyroscopic Pumping} \label{sec:two}
In this section, we briefly discuss a qualitative condition of the Reynolds stress for driving a double-cell meridional circulation.
Using a spherical geometry $(r,\theta,\phi)$, the velocity field $\bm{v}$ is written as,
\begin{eqnarray}
\bm{v} &=&\langle\bm{v}\rangle+\bm{v}',\\
\langle\bm{v}\rangle&=&\bm{v}_{m}+\bm{v}_{\phi} \nonumber \\
&=&v_{r}\bm{e}_{r}+v_{\theta}\bm{e}_{\theta}+r\sin{\theta}\Omega\bm{e}_{\phi},
\end{eqnarray}
where $\langle \  \rangle$ denotes zonal averaging. $\langle\bm{v}\rangle$ and $\bm{v}'$ express the mean velocity and the turbulence respectively, and $\Omega$ denotes the total angular velocity.
The mean $\langle\bm{v}\rangle$ represents the large-scale flows such as meridional circulation $\bm{v}_{m}=v_{r}\bm{e}_{r}+v_{\theta}\bm{e}_{\theta}$ and differential rotation $\bm{v}_{\phi}=r\sin{\theta}\Omega\bm{e}_{\phi}$.

Now, the equation of gyroscopic pumping can be expressed as,
\begin{eqnarray}
\rho_{0}\bm{v}_{m}\cdot\nabla\mathcal{L}=-\nabla\cdot
    (\rho_{0}r\sin{\theta}\langle\bm{v}_{m}'v_{\phi}'\rangle), \label{eq:gyro1}
\end{eqnarray}
when an anelastic approximation is valid for the mean flow, i.e., $\nabla\cdot(\rho_{0}\bm{v}_{m})=0$.
Here $\rho_0$ denotes the time-independent background density and $\mathcal{L} = r^{2}\sin^{2}{\theta}\Omega$ denotes the mean angular momentum density per unit mass. Equation (\ref{eq:gyro1}) describes that, in a stationary state, the angular momentum advected by the meridional flow balances with the angular momentum transported by $\phi$ components of the Reynolds stress.
Since, in general, $\nabla\mathcal{L}$ is directed almost cylindrically outward with respect to the rotational axis within the convection zone, it would be constructive to introduce a new coordinate $\zeta=r\sin{\theta}$ denoting the momentum arm.
Equation (\ref{eq:gyro1}) then reduces to,
\begin{eqnarray}
  \rho_{0}v_{\zeta}\frac{\partial\mathcal{L}}{\partial\zeta}\approx-\nabla\cdot
    (\rho_{0}r\sin{\theta}\langle\bm{v}_{m}'v_{\phi}'\rangle), \label{eq:gyro2}
\end{eqnarray}
where $v_{\zeta}$ is the cylindrically outward velocity.
\begin{table}[tb]
\begin{center}
\caption{Signs of each of the quantities in Eq.(\ref{eq:gyro2})}
\begin{tabular}{ccccc} \hline\hline
 Layer&& $v_{\zeta}$& $\nabla\cdot(\rho_{0}r\sin{\theta}\langle\bm{v}_{m}'v_{\phi}'\rangle)$ & $\langle v_{r}'v_{\phi}'\rangle$  \\ \hline 
upper&:& negative  &positive &  \\
middle&:& positive &negative & \raisebox{1.2 em}[0cm][0cm]{negative} \\ 
lower&:& negative &positive & \raisebox{0.55 em}[0cm][0cm]{positive} \\ \hline \label{table:1}
\end{tabular}
\end{center}
\end{table}

For double-cell type meridional circulation suggested by \citet{zhao2013}, $v_{\zeta}$ becomes negative both near the surface and the base of the convection zone where the meridional flow is directed poleward, whereas $v_{\zeta}$ takes positive value in the middle layer due to the equatorward return flow.
Since $\partial\mathcal{L}/\partial\zeta$ is generaly almost uniformly positive, the left hand side of the equation (\ref{eq:gyro2}) takes negative, positive, and negative values from lower, middle, to the upper convection zone, respectively.
Thus, $\langle\bm{v}_{m}'v_{\phi}'\rangle$ must be a radially converging vector into the middle convection zone and we can specify a Reynolds stress such that $\langle v_{r}'v_{\phi}'\rangle$ takes positive value in the lower half and becomes negative in the upper half of the convection zone as the simplest candidate which satisfies the condition above (Table \ref{table:1}). We can speculate, therefore, that the double-cell meridional circulation as observed might be maintained when the Reynolds stress transports the angular momentum radially upward (downward) in the lower (upper) convection zone. We quantitatively confirm this speculation by conducting a set of numerical simulations using a mean-field hydrodynamic model.

\section{Model} \label{sec:three}
We solve the hydrodynamic equations for a rotating system and investigate the large-scale flow structures, differential rotation and meridional circulation, using an axisymmetric mean-field model similar to that of \citet{rempel2005}.
In our mean-field model, all processes on the convective scale such as turbulent viscous dissipation, turbulent heat conduction, and turbulent angular momentum transport are parameterized and given explicitly in our basic equations.
Thus, the differential rotation and the meridional circulation are calculated self-consistently, which makes it possible for us to discuss the condition of the Reynolds stress for obtaining a double-cell meridional circulation along with the solar-like differential rotation.
We do not aim at a realistic simulation of the solar convection zone by this mean-field model but rather investigate the hydrodynamical balance for the double-cell meridional circulation.

We consider the perturbations with respect to the background reference state that is rigidly-rotating at an angular velocity of the radiative zone $\Omega_{0}/2\pi =430 \ \mathrm{nHz}$.
The basic assumptions used in our model are summarized as follows.
\begin{enumerate}
  \setlength{\parskip}{0cm}
  \setlength{\itemsep}{0cm}
\item Background stratification of the reference state is adiabatic and spherically symmetric.
\item There is no meridional flow in the reference state.
\item Energy balance is achieved in the reference state.
\item The perturbations of density and pressure associated with the differential rotation are small, i.e., $\rho_{0} \gg |\rho_{1}|$ and $p_{0} \gg |p_{1}|$. Here, we define quantities in the reference state with a subscript $0$, and perturbations from the reference state with a subscript $1$. We neglect the second-order terms of these perturbations.
\end{enumerate}

\subsection{Basic Equations} \label{subsec:basic equations}

The axisymmetric, fully compressible hydrodynamic equations in a rotating system can be expressed using a spherical geometry $(r,\theta,\phi)$ as,
\begin{eqnarray}
\frac{\partial \rho_{1}}{\partial t}&=&-\frac{1}{r^{2}}\frac{\partial}{\partial r}(r^{2}v_{r}\rho_{0})
    -\frac{1}{r\sin{\theta}}\frac{\partial}{\partial \theta}(\sin{\theta}v_{\theta}\rho_{0}), \label{eq:conti} \\
\frac{\partial v_{r}}{\partial t}&=&-v_{r}\frac{\partial v_{r}}{\partial r}-\frac{v_{\theta}}{r}
    \frac{\partial v_{r}}{\partial \theta}+\frac{v_{\theta}^{2}}{r}-\frac{1}{\rho_{0}}\left[\frac
      {\partial p_{1}}{\partial r}+\rho_{1} g\right] \nonumber \\
    &&+(2\Omega_{0}\Omega_{1}+\Omega_{1}^{2})r\sin^{2}
    {\theta}+\frac{F_{r}}{\rho_{0}}, \label{eq:motionr}\\
\frac{\partial v_{\theta}}{\partial t}&=&-v_{r}\frac{\partial v_{\theta}}{\partial r}-\frac{v_{\theta}}
    {r}\frac{\partial v_{\theta}}{\partial \theta}-\frac{v_{r}v_{\theta}}{r}-\frac{1}{\rho_{0} r}
    \frac{\partial p_{1}}{\partial \theta} \nonumber \\
    &&+(2\Omega_{0}\Omega_{1}+\Omega_{1}^{2})r\sin{\theta}
    \cos{\theta}+\frac{F_{\theta}}{\rho_{0}},\\
\frac{\partial \Omega_{1}}{\partial t}&=&-\frac{v_{\theta}}{r^{2}}\frac{\partial}{\partial r}
     [r^{2}(\Omega_{0}+\Omega_{1})] \nonumber \\
     &&-\frac{v_{\theta}}{r\sin^{2}{\theta}}\frac{\partial}{\partial\theta}
     [\sin^{2}{\theta}(\Omega_{0}+\Omega_{1})]+\frac{F_{\phi}}{\rho_{0} r\sin{\theta}}, \\
\frac{\partial s_{1}}{\partial t}&=&-v_{r}\frac{\partial s_{1}}{\partial r}-\frac{v_{\theta}}{r}\frac
     {\partial s_{1}}{\partial \theta}+v_{r}\frac{\gamma\delta}{H_{p}} \nonumber \\
     &&+\frac{\gamma-1}{p_{0}}Q+\frac{1}
    {\rho_{0} T_{0}}\nabla\cdot(\rho_{0} \kappa_{\mathrm{t}}T_{0}\nabla s_{1}), \label{eq:entropy}
\end{eqnarray}
where $(v_{r}$, $v_{\theta})$ denotes the meridional flow, $\Omega_{1}$ represents the relative angular velocity with respect to that of radiative zone $\Omega_{1}=\Omega-\Omega_{0}$, and $s_{1}$ is the dimensionless entropy perturbation normalized by the specific heat capacity at constant volume $c_{v}=(\gamma-1)^{-1}R/\mu$, with $\gamma=5/3$. $g$, $\delta$, and $\kappa_{\mathrm{t}}$ denote gravitational acceleration, superadiabaticity, and coefficient of turbulent thermal conductivity, respectively.
The pressure perturbation and pressure scale height are expressed as,
\begin{eqnarray}
  p_{1}&=&p_{0}\left( \gamma\frac{\rho_{1}}{\rho_{0}}+s_{1} \right), \\
  H_{p}&=&\frac{p_{0}}{\rho_{0}g}.
\end{eqnarray}

Since molecular viscosity is negligible in the sun's interior, the viscous force $\bm{F}$ can be expressed only by the Reynolds stress $R_{ik}$ as follows;
\begin{eqnarray}
F_{r}&=&\frac{1}{r^{2}}\frac{\partial}{\partial r}(r^{2}R_{rr})+\frac{1}{r\sin{\theta}}\frac{\partial}
{\partial \theta}(\sin{\theta}R_{\theta r}) \nonumber \\
&&-\frac{R_{\theta\theta}+R_{\phi\phi}}{r},\\
F_{\theta}&=&\frac{1}{r^{2}}\frac{\partial}{\partial r}(r^{2}R_{r\theta})+\frac{1}{r\sin{\theta}}
\frac{\partial}{\partial \theta}(\sin{\theta}R_{\theta \theta}) \nonumber \\
&&+\frac{R_{r\theta}-R_{\phi\phi}
    \cot{\theta}}{r},\\
F_{\phi}&=&\frac{1}{r^{2}}\frac{\partial}{\partial r}(r^{2}R_{r\phi})+\frac{1}{r\sin{\theta}}
\frac{\partial}{\partial \theta}(\sin{\theta}R_{\theta \phi}) \nonumber \\
&&+\frac{R_{r\phi}+R_{\theta\phi}
    \cot{\theta}}{r},
\end{eqnarray}
with
\begin{eqnarray}
R_{ik}=-\rho_{0}\langle v_{i}'v_{k}' \rangle.
\end{eqnarray}
$R_{ik}$ is assumed to be divided into a diffusive part and a non-diffusive part as follows,
\begin{eqnarray}
R_{ik}=\rho_{0}\left[\nu_{\mathrm{v}}\left(E_{ik}-\frac{2}{3}\delta_{ik}\nabla\cdot\bm{v}
    \right)+\nu_{\mathrm{l}}\Lambda_{ik}\Omega_{0}\right].
\end{eqnarray}
Here, $\nu_{\mathrm{v}}$ and $\nu_{\mathrm{l}}$ are the coefficients of turbulent diffusivity for viscous \replaced{dissipative}{diffusive} part and \replaced{non-dissipative}{non-diffusive} part \added{which does not contain any spatial derivatives of velocity field}, respectively.
$E_{ik}$ denotes the deformation tensor and is expressed in spherical coordinates by,
\begin{eqnarray}
E_{rr}&=&2\frac{\partial v_{r}}{\partial r},\\
E_{\theta\theta}&=&\frac{2}{r}\left(\frac{\partial v_{\theta}}{\partial\theta}+v_{r}\right),\\
E_{\phi\phi}&=&\frac{2}{r}(v_{r}+v_{\theta}\cot{\theta}),\\
E_{r\theta}&=&E_{\theta r}=r\frac{\partial}{\partial r}\left(\frac{v_{\theta}}{r}\right)
    +\frac{1}{r}\frac{\partial v_{r}}{\partial\theta},\\
E_{r\phi}&=&E_{\phi r}=r\sin{\theta}\frac{\partial\Omega_{1}}{\partial r},\\
E_{\theta\phi}&=&E_{\phi\theta}=\sin{\theta}\frac{\partial\Omega_{1}}{\partial\theta}.
\end{eqnarray}
The \replaced{non-dissipative}{non-diffusive} part of the Reynolds stress describes the effect of the turbulent angular momentum transport due to the anisotropic turbulence. Since the anisotropy is mainly attributed to the effect of the Coriolis forces related with the background rotation, the \replaced{non-dissipative}{non-diffusive} part is assumed to be proportional to $\Omega_{0}$ \citep[$\Lambda$-effect,][]{rudiger93,kichatinovrudiger93}.
In this study, we parameterize $\Lambda_{ik}$ in the non-diffusive part so that it satisfies the condition of the Reynolds stress needed for driving the double-cell meridional circulation derived in Section \ref{sec:two}.
Parameterization of $\Lambda_{ik}$  is described in detail in the later Section \ref{subsec:lambda-modeling}.

The amount of energy that is converted by the Reynolds stress from kinetic energy to internal energy is given by,
\begin{eqnarray}
Q=\sum_{i,k}\frac{1}{2}E_{ik}R_{ik}.
\end{eqnarray}

\subsection{Background Stratification}
For an adiabatic background stratification for the reference state, the same formulations presented in \citet{rempel2005} are adopted for density $\rho_{0}$, pressure $p_{0}$, temperature $T_{0}$, and gravitational acceleration $g$.
\begin{eqnarray}
\rho_{0}(r)&=&\rho_{\mathrm{bc}}\left[1+\frac{\gamma-1}{\gamma}\frac{r_{\mathrm{bc}}}{H_{\mathrm{bc}}}
    \left(\frac{r_{\mathrm{bc}}}{r}-1\right)\right]^{1/\gamma-1},\\
p_{0}(r)&=&p_{\mathrm{bc}}\left[1+\frac{\gamma-1}{\gamma}\frac{r_{\mathrm{bc}}}{H_{\mathrm{bc}}}
    \left(\frac{r_{\mathrm{bc}}}{r}-1\right)\right]^{\gamma/\gamma-1},\\
T_{0}(r)&=&T_{\mathrm{bc}}\left[1+\frac{\gamma-1}{\gamma}\frac{r_{\mathrm{bc}}}{H_{\mathrm{bc}}}
  \left(\frac{r_{\mathrm{bc}}}{r}-1\right)\right],\\
g(r)&=&g_{\mathrm{bc}}\left(\frac{r}{r_{\mathrm{bc}}}\right)^{-2},
\end{eqnarray}
where $\rho_{\mathrm{bc}}$, $p_{\mathrm{bc}}$, $T_{\mathrm{bc}}$, $g_{\mathrm{bc}}$ and $H_{\mathrm{bc}}=p_{\mathrm{bc}}/(\rho_{\mathrm{bc}}g_{\mathrm{bc}})$ are the values of density, pressure, temperature, and pressure scale height at the base of the convection zone $r_{\mathrm{bc}}=0.71 \ R_{\odot}$, with the solar radius $R_{\odot}=7\times 10^{10} \ \mathrm{cm}$.
We adopt solar values $\rho_{\mathrm{bc}}=0.2\ \mathrm{g}\ \mathrm{cm}^{-3}$, $p_{\mathrm{bc}}=6\times 10^{13}\ \mathrm{dyn}\ \mathrm{cm}^{-2}$, $T_{\mathrm{bc}}=mp_{\mathrm{bc}}/(k_{B}\rho_{\mathrm{bc}})\approx 1.82\times 10^{6}\ \mathrm{K}$, and $g_{\mathrm{bc}}=5.2\times 10^{4}\ \mathrm{cm}\ \mathrm{s}^{-2}$, where $k_{\mathrm{B}}$ is the Boltzman constant and $m$ is the mean particle mass.

It would be reasonable to assume such an adiabatic stratification for background quantities as long as we use a superadiabaticity defined by,
\begin{eqnarray}
  \delta&=&\nabla-\nabla_{\mathrm{ad}} \nonumber \\
  &=&-\frac{H_{p}}{\gamma}\frac{ds_{0}}{dr},
\end{eqnarray}
which is far smaller than the order of unity, i.e., $|\delta|\ll 1$.
We assume the following profile for superadiabaticity $\delta(r)$.
\begin{eqnarray}
\delta(r)=\frac{\delta_{\mathrm{os}}}{2}
      \left[1-\tanh\left(\frac{r-r_{\mathrm{tran}}}{d_{\mathrm{tran}}}\right)\right],
\end{eqnarray}
where $\delta_{\mathrm{os}}$ denotes the superadiabaticity value at the subadiabatic layer below the convection zone.
Superadiabaticity smoothly connects with the adiabatic profile in the bulk of the convection zone at the transition radius $r_{\mathrm{tran}}$ with the steepness of the transition $d_{\mathrm{tran}}$. Here, we adopt $\delta_{\mathrm{os}}=-1.5\times 10^{-5}$, $r_{\mathrm{tran}}=0.725\ R_{\odot}$, and $d_{\mathrm{tran}}=0.0125 \ R_{\odot}$ in all of our calculations.

In our model, inclusion of subadiabatic layer below the base of the convection zone produces the entropy perturbation through interactions with the radial meridional flow (third term in equation (\ref{eq:entropy})), which is proposed by \citet{rempel2005} as a possible source for the latitudinal entropy gradient that is needed for thermal wind balance of the non-Taylor Proudman differential rotation \citep{miesch2006}.
In fact, stratification becomes superadiabatic in the upper convection zone in our real sun \citep{skaley1991}.
Different profiles of $\delta$ with such a superadiabatic convection zone may primarily affect the Taylor-Proudman balance of the differential rotation but the influence on the meridional circulation, which is of our primary interest in this paper, is very weak \citep{rempel2005}.

We must also note that there are other possible theories on the generation mechanism of the entropy perturbation. \citet{kitchatinov1995} proposed that an latitudinal temperature gradient is produced by an anisotropic energy transport, while \citet{masada2011} suggested that the turbulence induced by magnetorotational instability plays an important role in generating entropy near the higher latitude tachocline via the exceptional turbulent heating.
The Taylor-Proudman balance of the differential rotation may change when these other entropy sources are considered.
We will discuss the Taylor-Proudman balance in our model in the later Section \ref{subsec:TPbalance}.

\subsection{Turbulent Diffusivity Profiles}
We assume constant values for turbulent viscosity and thermal conductivity within the convection zone and the radiative zone with a thin transition layer.
We assume the following diffusivity profiles having only radial dependence,
\begin{eqnarray}
\nu_{\mathrm{v}}(r)&=&\nu_{\mathrm{os}}+(\nu_{0}-\nu_{\mathrm{os}}) f_{\mathrm{tran}}(r) f_{\mathrm{c}}(r),\\
\nu_{\mathrm{l}}(r)&=&\nu_{0} f_{\mathrm{tran}}(r)  f_{\mathrm{c}}(r),\\
\kappa_{\mathrm{t}}(r)&=&\kappa_{\mathrm{os}}+(\kappa_{0}-\kappa_{\mathrm{os}}) f_{\mathrm{tran}}(r)  f_{\mathrm{c}}(r),
\end{eqnarray}
with
\begin{eqnarray}
    f_{\mathrm{tran}}(r)&=&\frac{1}{2}\left[1+\tanh\left(\frac{r-r_{\mathrm{tran}}+\Delta}{d_{\kappa\nu}}\right)\right], \\
  \Delta&=&d_{\kappa\nu}\tanh^{-1}(2\alpha_{\kappa\nu}-1), \\
  f_{\mathrm{c}}(r)&=&\frac{1}{2}\left[1+\tanh\left(\frac{r-r_{\mathrm{bc}}}{d_{\mathrm{bc}}}\right)\right],
\end{eqnarray}
where $\nu_{0}$ and $\kappa_{0}$ are the values of turbulent diffusivities within the convection zone and $\nu_{\mathrm{os}}$ and $\kappa_{\mathrm{os}}$ are those within the subadiabatic layer. 
Note that we impose moderate diffusivities below the convection zone in order to achieve a shear layer near the uniformly-rotating radiative core, and in addition, to make it possible for the entropy perturbation to spread out into the bulk of the convection zone.
Here, we adopt $\nu_{0}=\kappa_{0}=3.0\times 10^{12}\ \mathrm{cm}^{2}\ \mathrm{s}^{-1}$ and set the viscosity and thermal conductivity in the subadiabatic layer $\nu_{\mathrm{os}}$, $\kappa_{\mathrm{os}}$ to $2 \%$ and $0.2 \%$ relative to those within the convection zone, respectively.
The radial step function $f_{\mathrm{tran}}(r)$ smoothly connects the diffusivity values in the subadiabatic layer with those in the upper convection zone, in which the parameter $\alpha_{\kappa\nu}=f_{\mathrm{tran}}(r_{\mathrm{tran}})$ determines diffusivity values at $r=r_{\mathrm{tran}}$.
The other function $f_{\mathrm{c}}(r)$ ensures that the diffusivities drop significantly towards the radiative interior.
$d_{\kappa\nu}$ and $d_{\mathrm{bc}}$ denote widths of these transition layers.
In all our simulations, we use the values, $\alpha_{\kappa\nu}=0.1$, $d_{\kappa\nu}=0.025 \ R_{\odot}$, and $d_{\mathrm{bc}}=0.0125\ R_{\odot}$.

\subsection{Parameterization of the $\Lambda$-Effect} \label{subsec:lambda-modeling}
\begin{figure}[t!]
\figurenum{1}
\plotone{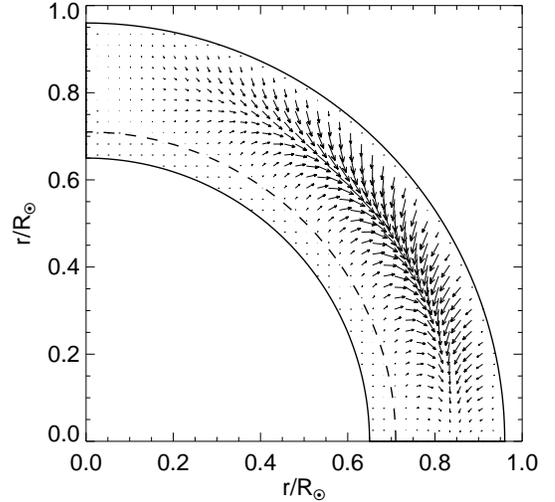}
\caption{\replaced{Non-dissipative}{Non-diffusive} part the Reynolds stress ($\Lambda$-effect) is illustrated. $-\nu_{\mathrm{l}}\bm{\Lambda}$ which specifies the direction and the amplitude of the turbulent angular momentum flux is plotted on the meridional plane for our reference case $C_{\mathrm{l}}=0.3$. Dashed line indicates the base of the convection zone, $r_{\mathrm{bc}}=0.71 \ R_{\odot}$.} \label{fig:lambda}
\end{figure}
In this Section, non-diffusive part of the Reynolds stress tensor, $\Lambda_{ik}$, is specified \added{such that the condition of the Reynolds stress for the double-cell meridional circulation discussed in Section \ref{sec:two} is satisfied by the non-diffusive part}.
In our study, we do not consider the turbulent momentum transport within the meridional plane, and thus, $\Lambda_{ik}$ is assumed to have non-zero components only when $k$ takes $\phi$.
Here, let us define a dimensionless vector on a meridional plane that specifies the amplitude and the direction of the $\Lambda$-effect,
\begin{eqnarray}
\bm{\Lambda}=\Lambda_{r\phi}\bm{e}_{r}+\Lambda_{\theta\phi}\bm{e}_{\theta}.
\end{eqnarray}
Now, the angular momentum flux is divided into the turbulent diffusive flux and the turbulent angular momentum flux using $\bm{\Lambda}$ as,
\begin{eqnarray}
  \rho_{0}r\sin{\theta}\langle \bm{v}_{m}' v_{\phi}' \rangle &=&
    - \left( \rho_{0}\nu_{\mathrm{v}}r^{2}\sin^{2}{\theta}\nabla\Omega_{1} \right. \nonumber \\
           && \ \ \ \ \ \ \ \ + \left. \rho_{0}\nu_{\mathrm{l}}r\sin{\theta} \bm{\Lambda} \Omega_{0} \right).
\end{eqnarray}
Note that, by definition, $\bm{\Lambda}$ is anti-parallel to the turbulent angular momentum flux.

We set the $\Lambda$-effect expressed in the following form,
\begin{eqnarray}
\bm{\Lambda}
= \Lambda_{0}\tilde{f}(r,\theta)\left(\begin{array}{c}
                            \cos{(\theta +\lambda(r,\theta))}\\
                            -\sin{(\theta +\lambda(r,\theta))}\\
                            \end{array}\right) h(r), \label{eq:lambda_tensor}
\end{eqnarray}
where $\Lambda_{0}$ is a dimensionless parameter with the order of unity that specifies the overall amplitude of the $\Lambda$-effect and $\lambda$ is the inclination of $\bm{\Lambda}$ with respect to the rotational axis.
Spatial dependence is described by a distribution function $\tilde{f}(r,\theta)$ as, 
\begin{eqnarray}
\tilde{f}(r,\theta)&=&\frac{f(r,\theta)}{\mathrm{max} |f(r,\theta)|},\\
f(r,\theta)&=&\sin^{2}{\theta}\cos{\theta}\tanh\left(\frac{r_{\mathrm{max}}-r}{d_{\mathrm{sf}}}\right). \label{eq:f-lambda}
\end{eqnarray}
Here, $\sin^{2}{\theta}$ is multiplied in order to avoid the divergence of the Reynolds stress near the pole. $\cos{\theta}$ satisfies the antisymmetric boundary condition at the equator.
The hyperbolic-tangent part ensures that the amplitude of turbulent angular momentum flux drops significantly towards the surface.\deleted{where the turbulence is likely to become isotropic and the effect of the Coriolis force is likely to diminish.}We set a value $d_{\mathrm{sf}}=0.025 \ R_{\odot}$ for this transition width.

Considering the condition of the Reynolds stress \replaced{discussed in Section \ref{sec:two}}{that the angular momentum should be transported radially inward (outward) in the upper (lower) half of the convection zone}, we treat the direction and the amplitude of the turbulent angular momentum flux as functions of space that are specified by $\lambda(r,\theta)$ and $h(r)$, respectively.
\begin{eqnarray}
  \lambda(r,\theta)&=&\lambda_{\mathrm{l}}(\theta)+\frac{\lambda_{\mathrm{u}}(\theta)-\lambda_{\mathrm{l}}(\theta)}{2} \nonumber\\
  && \ \ \ \ \ \ \ \ \ \times\left[1+\tanh\left(\frac{r-r_{\mathrm{mid}}}{d_{\mathrm{mid}}}\right)\right],\\
h(r)&=&C_{\mathrm{l}}+\frac{1-C_{\mathrm{l}}}{2}\left[1+\tanh\left(\frac{r-r_{\mathrm{mid}}}{d_{\mathrm{amp}}}\right)\right]. \label{eq:hr}
\end{eqnarray}
Here, $\lambda_{\mathrm{l}}$ denotes the inclination of $\bm{\Lambda}$ in the lower part of the convection zone $r<r_{\mathrm{mid}} = 0.825 \ R_{\odot}$ and $\lambda_{\mathrm{u}}$ for the upper part $r_{\mathrm{mid}}<r$.
For smoothly connecting the turbulent angular momentum fluxes with different directions, transition layers with the thickness $d_{\mathrm{mid,amp}}$ are inserted.
Transition widths are specified as $d_{\mathrm{mid}}=0.085 \ R_{\odot}$ and $d_{\mathrm{amp}} = 0.075 \ R_{\odot}$.
$C_{\mathrm{l}}$ appeared in the radial function $h(r)$ specifies the relative amplitude of the $\Lambda$-effect in the lower part of the convection zone with respect to the upper layer.
$C_{\mathrm{l}}$ is treated as a free parameter along with $\Lambda_{0}$ in this study.
We specify the directions $\lambda_{\mathrm{l}}$ and $\lambda_{\mathrm{u}}$ as follows,
\begin{eqnarray}
  \lambda_{\mathrm{l}}&=&165^{\circ} ,\\
  \lambda_{\mathrm{u}}&=& \lambda_{\mathrm{eq}}+\frac{\lambda_{\mathrm{pl}}-\lambda_{\mathrm{eq}}}{2}\left[1+\tanh\left(\frac{\theta-\theta_{\mathrm{mid}}}{\Pi_{\mathrm{mid}}}\right) \right].
\end{eqnarray}
Here, we set $\theta_{\mathrm{mid}}=45^{\circ}$, $\Pi_{\mathrm{mid}}=15^{\circ}$, $\lambda_{\mathrm{eq}}=15^{\circ}$, and $\lambda_{\mathrm{pl}}=-\theta$.
The turbulent angular momentum flux is directed upward with the inclination $15 ^{\circ}$ to the rotational axis in the lower convection zone.
In the upper convection zone, on the other hand, the flux is latitudinally-directed in high latitudes and changes its direction to radially inward near the equator as shown in Figure \ref{fig:lambda}.

\added{
\citet{kichatinovrudiger93} investigated the forms of $\Lambda_{r\phi}$ and $\Lambda_{\theta\phi}$, especially their dependence on the radial distance, latitude, and rotation rate, using a quasilinear theory.
One of their main results that the horizontal component $\Lambda_{\theta\phi}$ is negative in the northern hemisphere, having $\propto\sin^{2}{\theta}\cos{\theta}$ dependence, is implemented in our model. This negative $\Lambda_{\theta\phi}$ or the resulting positive horizontal correlation $\langle v_{\theta}'v_{\phi}'\rangle$, which is responsible for the observed equatorial acceleration, has been repeatedly confirmed by both local convection simulations \citep{chan2001,kapyla2004,rudiger2005} and global spherical simulations \citep{brun2004,miesch2008,kapyla2011,kapyla2014}.}

\added{
On the other hand, the functional form of the vertical component $\Lambda_{r\phi}$ developed by \citet{kichatinovrudiger93} turns out to be unable to explain some of the results obtained by direct numerical simulations, and thus, $\Lambda_{r\phi}$ is parameterized only by considering the condition discussed in Section \ref{sec:two} without resorting to the quasilinear theory in our model.
Still, it would be noteworthy that some direct numerical simulations provide us a good reason to set the turbulent angular momentum flux upward (downward) in the lower (upper) convection zone as illustrated in Figure \ref{fig:lambda}.
Conducting local convection simulations under the different rotational influences, \citet{kapyla2004} found that the radial-azimuthal correlation $\langle v_{r}'v_{\phi}'\rangle$ shows a sign change from negative for slow rotation case ($Ro \gg 1$) to positive for fast rotation case ($Ro\ll 1$), where the Rossby number $Ro=v'/(2\Omega_{0}H_{p})$ is defined with convective velocity $v'$ and pressure scale height $H_{p}$.
If we assume that the viscous diffusive part of the Reynolds stress has just a negligible effect, the corresponding turbulent angular momentum flux is directed radially outward in low-Rossby regime and inward in high-Rossby regime, respectively. 
Since $v'$ and $H_{p}$ becomes larger and shorter respectively towards the surface according to global convection simulations, $Ro$ is expected to be small at the base and become larger near the surface, and thus, the angular momentum is expected to be transported outward (inward) in the lower (upper) convection zone.
The amplitude ratio of the $\Lambda$-effect between lower and upper convection zone $C_{\mathrm{l}}$ should reflect the radial dependence of $Ro$, which is difficult to determine by the current global convection simulations where $v'$ is likely to be overestimated \citep{miesch2008,hanasoge2012}, and thus, is treated as a free parameter in this paper.
More detailed discussions on the validity of our parameterization of the $\Lambda$-effect will be addressed again in later Section \ref{sec:validity} using the computed Reynolds stress. 
}

\subsection{Numerical Methods}
We solve the equations (\ref{eq:conti})$-$(\ref{eq:entropy}) numerically for the northern hemisphere of the meridional plane, using the second-order centered-differencing method for space and the fourth-order Runge-Kutta scheme for time integration \citep{vogler2005}.
We adopt the same artificial viscosity as \citet{rempel2014} which is added on all variables. 
The numerical domain extends from $r_{\mathrm{min}}=0.65 \ R_{\odot}$ to $r_{\mathrm{max}}=0.96 \ R_{\odot}$ in radius.
We use a uniform resolution of $70$ points for both radial and latitudinal directions.
Each numerical calculation is run until the large-scale flows become stationary.

At the initial state, we set all the variables $\rho_{1}$, $v_{r}$, $v_{\theta}$, $\Omega_{1}$, and $s_{1}$ to zero.
At the surface $r=r_{\mathrm{max}}$, we impose an impenetrable boundary condition for $v_{r}$ and stress-free boundary conditions for $v_{\theta}$ and $\Omega_{1}$ $(R_{r\theta}=R_{r\phi}=0)$. We set the derivative of $s_{1}$ to zero:
\begin{eqnarray}
v_{r}=0,\\
\frac{\partial}{\partial r}\left( \frac{v_{\theta}}{r}\right)=0,\\
\frac{\partial\Omega_{1}}{\partial r}=0,\\
\frac{\partial s_{1}}{\partial r}=0.
\end{eqnarray}
 We set $\rho_{1}$ to make the right-hand side of the equation (\ref{eq:motionr}) equal to zero at the top boundary.
Except for $\Omega_{1}$, the same radial boundary conditions are imposed at the lower boundary. For $\Omega_{1}$, we impose a uniform rotating boundary condition $\Omega_{1}=0$.
At the pole and the equator, we set the boundary conditions considering an appropriate symmetricity;
\begin{eqnarray}
\frac{\partial \rho_{1}}{\partial\theta}=0,\\
\frac{\partial v_{r}}{\partial\theta}=0,\\
v_{\theta}=0,\\
\frac{\partial \Omega_{1}}{\partial\theta}=0,\\
\frac{\partial s_{1}}{\partial\theta}=0.
\end{eqnarray}

In order to relax the severe CFL condition of the sound wave mode, we use the Reduced Speed of Sound Technique \citep{rempel2005,hotta2014,hotta2015} and replace the equation of continuity (\ref{eq:conti}) by,
\begin{eqnarray}
\frac{\partial\rho_{1}}{\partial t}+\frac{1}{\xi^{2}}\nabla\cdot(\rho_{0}\bm{v}_{m})=0.
\end{eqnarray}
In this study, we adopt $\xi=100$ in all our calculations.

\section{Results} \label{sec:four}

\begin{table}[t]
 \begin{center} 
\caption{Parameters used and calculated in our model}
\begin{tabular}{cccccc} \hline\hline
  \renewcommand{\arraystretch}{1.8}
  Case&$C_{\mathrm{l}}$&  $\Lambda_{0}$ && $P_{\mathrm{nTP}}$ & $P_{\mathrm{cw}}$
  \renewcommand{\arraystretch}{1} \\ \hline
 1.........  &0.1 & 1.0  & . . .  &$5.19\times 10^{-2}$&$-2.87 \times 10^{-1}$ \\
 2.........  &0.3 & 0.8  & . . .  &$4.80\times 10^{-2}$&$-1.15 \times 10^{-1}$\\ 
 3.........  &0.6 & 0.6  & . . .  &$4.39\times 10^{-2}$&$4.32 \times 10^{-2}$\\ 
 4.........  &1.0 & 0.45  & . . .  &$4.08\times 10^{-2}$&$1.57 \times 10^{-1}$ \\
 \hline
\end{tabular}
 \end{center}
\vspace{-1.3\baselineskip}
\tablecomments{ $C_{\mathrm{l}}$ and $\Lambda_{0}$ are free parameters in our model specified for each case. $P_{\mathrm{nTP}}$ and $P_{\mathrm{cw}}$ are the parameters calculated to evaluate the extent of the deviation of the differential rotation from the Taylor-Proudman state (equation (\ref{eq:NTP})) and the strength of the lower clockwise meridional circulation cell (equation (\ref{eq:CW})), respectively.} \label{table:2}
    \end{table}


We run simulations for four cases, with Table \ref{table:2} showing the parameters for each case.
Starting from the zero initial condition, the dynamical balance between differential rotation and meridional circulation is established in about $15 \ \mathrm{yr}$, which can be defined as a dynamical time-scale $\tau_{\mathrm{dyn}}$.
After the stationary velocity field has achieved, there appears a secular change in entropy perturbation which proceeds on a thermal diffusive time-scale $\tau_{\mathrm{diff}}\approx 500 \ \mathrm{yr}$.
Since the aim of this paper is to discuss the angular momentum balance for the double-cell meridional circulation, we do not consider the physical processes working on a thermal diffusive time-scale for the following discussion.

In Section \ref{subsec:drmc}, we present a representative reference solution, computed with the parameter values $C_{\mathrm{l}}=0.3$ and $\Lambda_{0}=0.8$ (Case 2).
The influence on the Taylor-Proudman balance is addressed in Section \ref{subsec:TPbalance}.
Parameter dependence is discussed at last in Section \ref{subsec:parameter}.

\subsection{Reference Solution} \label{subsec:drmc}

\begin{figure*}[]
\figurenum{2}
\plotone{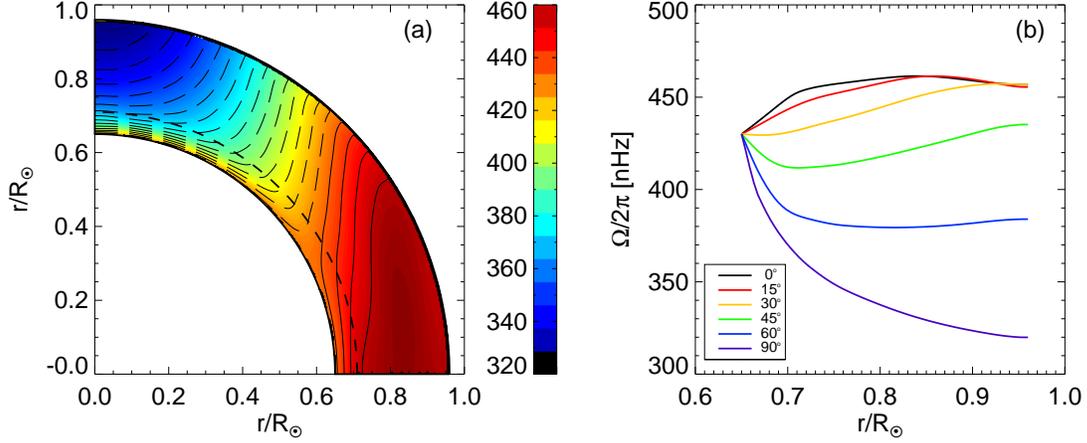}
\caption{Profile of differential rotation $\Omega/2\pi$ nHz in a dynamical stationary state ($t=\tau_{\mathrm{dyn}}$) for case 2. (a) The angular velocity distribution on the meridional plane. (b) The angular velocity at different colatitudes as functions of radial distance.} \label{fig:dr}
\end{figure*}

\begin{figure*}[]
\figurenum{3}
\plotone{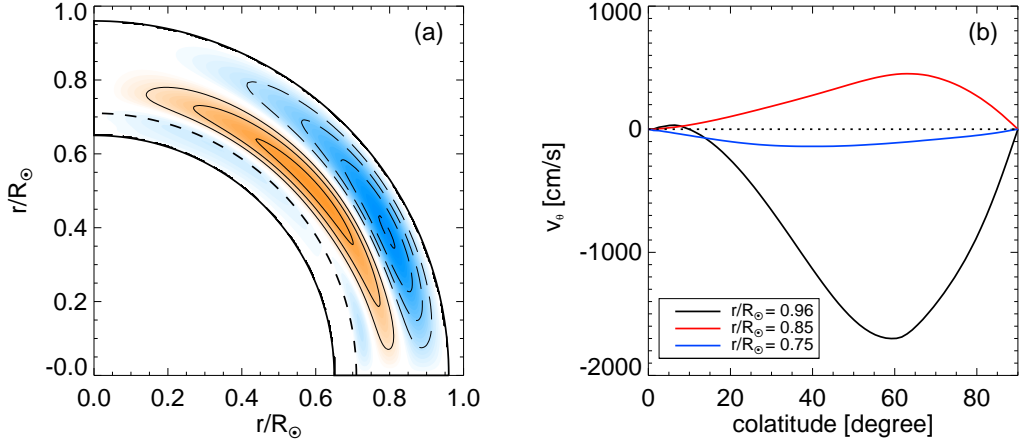}
\caption{Profile of meridional circulation $\bm{v}_{m}$ in a dynamical stationary state for case 2. (a) The stream-function $\Psi$, defined by $\rho_{0}\bm{v}_{m}=\nabla\times(\Psi \bm{e}_{\phi})$, is plotted on the meridional plane. Solid lines indicate streamlines of clockwise circulation flows and dashed lines for counterclockwise flows. (b) The latitudinal meridional velocity $v_{\theta}$ for different radii are plotted as functions of colatitude.} \label{fig:mc}
\end{figure*}

\begin{figure*}[]
\figurenum{4}
\plotone{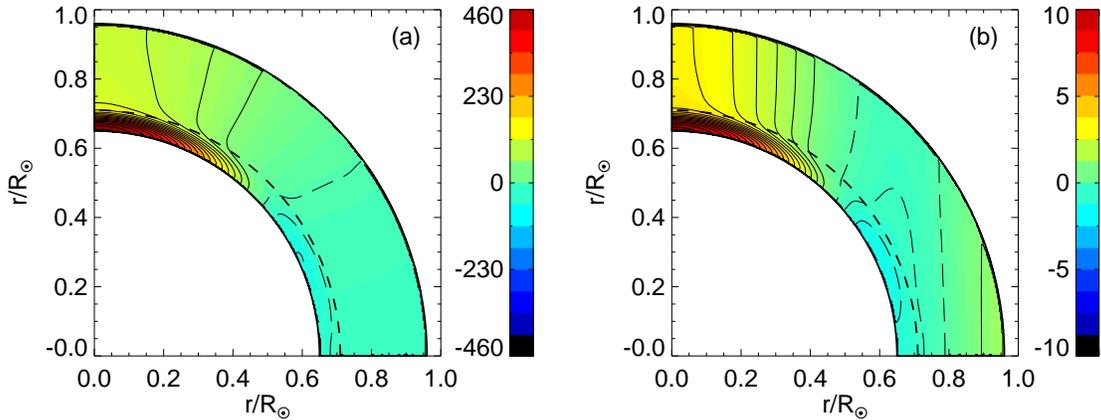}
\caption{Profiles of (a) entropy perturbation $c_{v}s_{1}$ in units of $\mathrm{erg}\ \mathrm{g}^{-1}\ \mathrm{K}^{-1}$ and (b) temperature perturbation $T_{1}=T_{0}[s_{1}+(\gamma -1)p_{1}/p_{0}]/\gamma$ in unit of $\mathrm{K}$ at $t=\tau_{\mathrm{dyn}}$ are plotted on meridional planes.} \label{fig:enttemp}
\end{figure*}

Figures \ref{fig:dr} and \ref{fig:mc} show the profiles of differential rotation and meridional circulation for case 2 respectively. As for differential rotation, the equator rotates at $\Omega/2\pi \approx 460 \ \mathrm{nHz}$ which is consistent with the helioseismic observations and the rotation rate almost monotonically decreases as latitude increases, as like our sun.
In our model, rigidly-rotating lower boundary, in conjunction with a finite value of $\nu_{\mathrm{v}}$, makes a strong angular velocity shear layer at the base of the convection zone, consisting a tachocline.
We can confirm that the contour lines of the angular velocity are conical especially in middle to high latitudes. This means that the Taylor-Proudman's theorem is broken by the negative latitudinal entropy gradient within the convection zone \citep{miesch2006} as shown in Figure \ref{fig:enttemp}(a). The temperature perturbation associated with this entropy perturbation shows a latitudinal difference of about $12\ \mathrm{K}$ at the base and about $5 \ \mathrm{K}$ throughout the convection zone as shown in Figure \ref{fig:enttemp}(b).

Nevertheless, there are several differences against the profile deduced by helioseismology. One major difference is the absence of the near surface shear layer \citep{howe2011}.
\added{In fact, some mean-field models have successfully reproduced the surface shear layer with non-vanishing inward angular momentum fluxes at the surface \citep{kitchatinov2005,rempel2005,kitchatinov2011}. \citet{rempel2005} presented two meridional flow profiles computed with and without the near surface shear layer and found almost no influence on the meridional flow structure except for the increase of poleward flow speed at the surface. Furthermore, the meridional force balance resulting from the correlation $\langle v_{r}'v_{\theta}'\rangle$, which is not parameterized in our model, is claimed to be crucial for the maintenance mechanism of the surface shear layer \citep{miesch2011} and also confirmed by the global simulation \citep{hotta2015}. For the above reasons, we exclude the near surface shear layer in our model by confining the computational domain below the subsurface region $r<0.96 \ R_{\odot}$, expecting that the impact on the meridional flow structure is very limited.}
\deleted{but, since our mean-field model cannot resolve the small-scale convections near the surface where the pressure scale height rapidly decreases, we do not discuss the effect of the near surface shear layer in this paper.}
Another difference is that the rotation rate near the equator shows a gradual decrease towards the surface. This comes from the fact that, in our parameterization of the $\Lambda$-effect, the angular momentum is transported radially inward in the upper half of the convection zone.
\added{Moreover, the tachocline is mostly confined beneath the base of the convection zone for the parameters used in our model, which is different from helioseismic inversions where the tachocline is overlapped with the convection zone \citep{howe2009}.}
The difference in the Taylor-Proudman balance is addressed in the next Section \ref{subsec:TPbalance}.

As for meridional circulation, we can confirm that the double-cell structure is achieved with the clockwise cell in the lower half and the counterclockwise cell in the upper half of the convection zone, which is similar to \citet{zhao2013}'s observational result.  
Due to the effect of the subadiabatic layer, the meridional flow is mostly excluded below the base of the convection zone $r<r_{\mathrm{bc}}$. The latitudinal velocity takes a maximum poleward speed $v_{\theta}=1700\ \mathrm{cm}\ \mathrm{s}^{-1} $ at the surface $r=0.96 \ R_{\odot}$ and reverses its direction to equatorward near $r=0.90\ R_{\odot}$. The equatorward velocity takes its maximum value $v_{\theta}=450\ \mathrm{cm} \ \mathrm{s}^{-1}$ at $r=0.85\ R_{\odot}$ and changes its direction again to poleward around $r=0.79\ R_{\odot}$. The poleward flow near the base of the convection zone is less than $200 \ \mathrm{cm} \ \mathrm{s}^{-1}$ as shown in Figure \ref{fig:mc}(b).
\added{Although the meridional flow amplitude presented in \citet{zhao2013} is larger by a factor of 5 especially in the lower convection zone than that of our reference model, it is also reported that the inversion with mass-conservation constraint reduces the flow amplitude in the deeper layer while keeping the basic double-cell profile. Our result is thus consistent with this statement.}

The mechanism of how the double-cell meridional circulation is driven along with the solar-like differential rotation in our model can be understood as follows. The $\phi$ component of the vorticity equation is expressed by,
\begin{eqnarray}
\frac{\partial\omega_{\phi}}{\partial t}=[\ .\ .\ .\ ]+r\sin{\theta}\frac{\partial\Omega^{2}}{\partial z}
    -\frac{g}{\gamma r}\frac{\partial s_{1}}{\partial\theta}, \label{eq:vort}
\end{eqnarray}
where $\omega_{\phi}=(\nabla\times\bm{v}_{m})_{\phi}$ is the meridional vorticity and $z$ denotes the rotational axis.
Here, for simplicity, advective and diffusive terms are neglected. Since the Reynolds stress takes the angular momentum away from the lower and upper parts of the convection zone and accumulate it into the middle layer (consider the divergence of the non-diffusive part of the Reynolds stress), $\partial\Omega^{2}/\partial z$ becomes negative near or below the base of the convection zone $r \lesssim r_{\mathrm{bc}}$, positive in low to middle convection zone $r\approx 0.8 \ R_{\odot}$, and again negative in the upper part $r\approx 0.9 \ R_{\odot}$.
Coriolis forces acting on these negative, positive, and negative $\partial\Omega^{2}/\partial z$, therefore, drive the counterclockwise, clockwise, and counterclockwise circulation cells, respectively, as expressed by the left hand side and the second term on the right hand side of the equation (\ref{eq:vort}). The counterclockwise circulation cell near or below the base of the convection zone produces positive (negative) entropy perturbation in high (low) latitudes due to the penetration of negative (positive) radial velocity into the subadiabatic region (through the third term in equation (\ref{eq:entropy})).
The buoyancy force arising from this entropy perturbation almost excludes the lowermost meridional circulation cell and the double-cell structure is obtained as a result.
The generated entropy perturbation below the convection zone can gradually spread out into the bulk of the convection zone due to the finite value of turbulent thermal conductivity $\kappa_{\mathrm{os}}$ in the overshoot region so that the negative latitudinal entropy gradient $\partial s_{1}/\partial \theta$ is established.
At the same time, solar-like differential rotation is driven by the latitudinal angular momentum transport by the Reynolds stress.
Differential rotation is finally balanced by the latitudinal entropy gradient such that the right hand side of the equation (\ref{eq:vort}) vanishes.
In other words, differential rotation is in a ``thermal wind balance'' in a stationary state: If significant latitudinal entropy gradients are present, non-Taylor Proudman differential rotation is achieved.  

\subsection{Taylor-Proudman Balance} \label{subsec:TPbalance}
\begin{figure*}[]
\figurenum{5}
\plotone{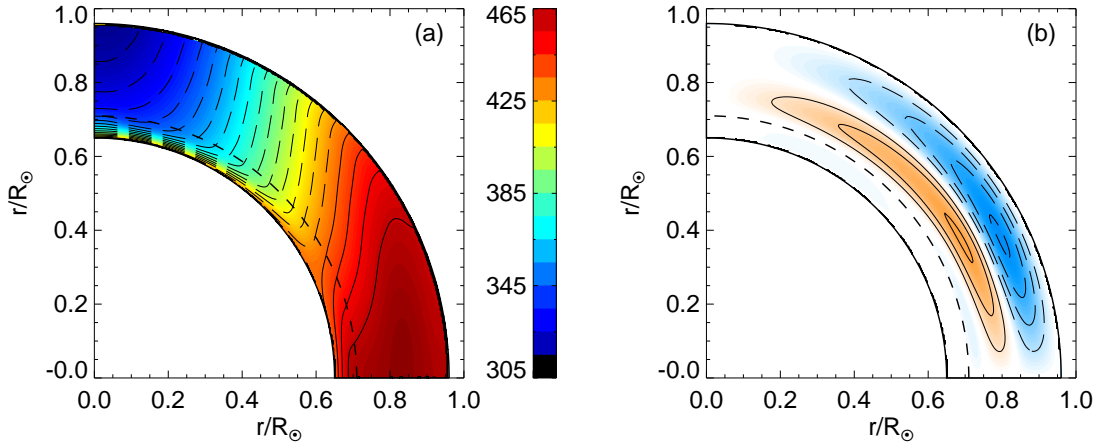}
\caption{Profiles of (a) differential rotation $\Omega/2\pi$ nHz and (b) meridional circulation $\bm{v}_{m}$ in a dynamical stationary state computed with a strongly subadiabatic layer below the base of the convection zone: The absolute value of the superadiabaticity is increased by a factor of 3, $\delta_{\mathrm{os}}=-4.5\times10^{-5}$.} \label{fig:deltaos}
\end{figure*}

Compared with the profile of the differential rotation presented in \cite{rempel2005} which is attained with the single-cell meridional circulation, the differential rotation in our model becomes close to the Taylor-Proudman state with cylindrical contour lines in low to middle latitudes (Figure \ref{fig:dr}(a)).
Figure \ref{fig:enttemp}(a) shows that the latitudinal entropy gradient in this region is very weak: $c_{v}[s_{1}(r_{\mathrm{max}},45^{\circ})-s_{1}(r_{\mathrm{max}},90^{\circ})]=35.8 \  \mathrm{erg} \ \mathrm{g}^{-1} \ \mathrm{K}^{-1}$.
Thus, the effect of the thermal wind becomes small in our double-cell case.

This can be explained as follows.
In the case of double-cell meridional circulation, radially inward meridional flow of the lower clockwise cell near the equator has the advective effect of suppressing the entropy perturbation from spreading into the bulk of the convection zone, leading to weaker latitudinal entropy gradient in low to middle latitudes. 
In a single-cell case, in contrast, the radially outward meridional flow near the base of the convection zone can help entropy perturbation to spread upward and to produce significant latitudinal entropy gradient in this region, by which the rotation profile can largely deviates from the Taylor-Proudman state.
It is, therefore, suggested that in our mean-field model the double-cell structure of the meridional circulation has an unfavorable effect on the realization of the non Taylor-Proudman differential rotation that is deduced by helioseismology compared with the single-cell case, mainly due to the advection of the entropy perturbation by the lower clockwise circulation cell near the equator.

It should be noted that our mean-field model has many important parameters, in addition to the $\Lambda$-effect, that can also affect the results. Therefore, realizing the non-cylindrical rotation profile similar to the helioseismic measurement along with the double-cell meridional circulation would not be a difficult task for our mean-field model.
For example, since the amplitude of the entropy \replaced{gradient}{perturbation} must depend on the background entropy stratification $\partial s_{0}/\partial r$, decreasing the superadiabaticity below the base of the convection zone $\delta_{os}$ \replaced{would}{can} retrieve the non Taylor-Proudman rotation profile while keeping the meridional flow structure almost unchanged. \added{Figure \ref{fig:deltaos} shows the result for a value of $\delta_{\mathrm{os}}$ 3 times larger than that of our reference case with all the other parameters kept identical. In this case, the increase in the entropy perturbation originated from the subadiabatic layer can almost be balanced by the increase in the differential rotation. We present this solution as an example to show that it is possible to obtain non-cylindrical rotational profile with our parameterization of the $\Lambda$-effect.}
However, since the aim here is to discuss the influence of the double-cell meridional circulation structure on the Taylor-Proudman balance by comparing with the single-cell case, we conduct \added{the rest of} our simulations with \replaced{identical}{the same} parameterizations for superadiabaticity $\delta_{os}$, turbulent thermal diffusivity $\kappa_{\mathrm{t}}$, and $\alpha_{\kappa\nu}$ (value of the diffusivities at $r_{\mathrm{tran}}$) as presented in \citet{rempel2005}'s single-cell case.

\subsection{Dependence on $\Lambda$-Effect} \label{subsec:parameter}
\begin{figure}[t]
\figurenum{6}
\plotone{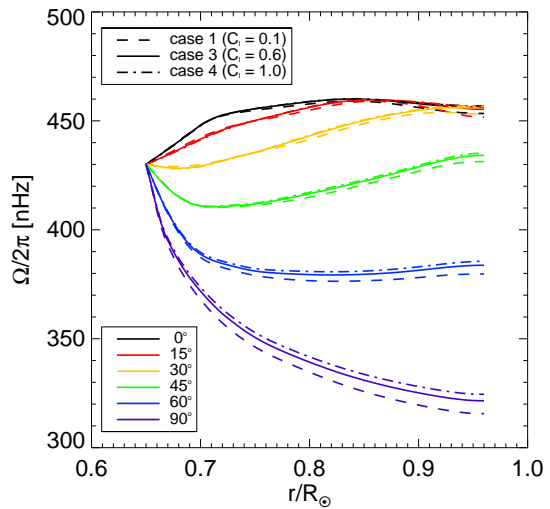}
\caption{Profiles of the differential rotation as functions of radial distance at different colatitudes for case 1 ($C_{\mathrm{l}} =0.1$), case 3 ($C_{\mathrm{l}}=0.6$), and case 4 ($C_{\mathrm{l}}=1.0$) with dashed lines, solid lines, and dash-dot lines, respectively.} \label{fig:drparam}
\end{figure}

\begin{figure}[!t]
\figurenum{7}
\plotone{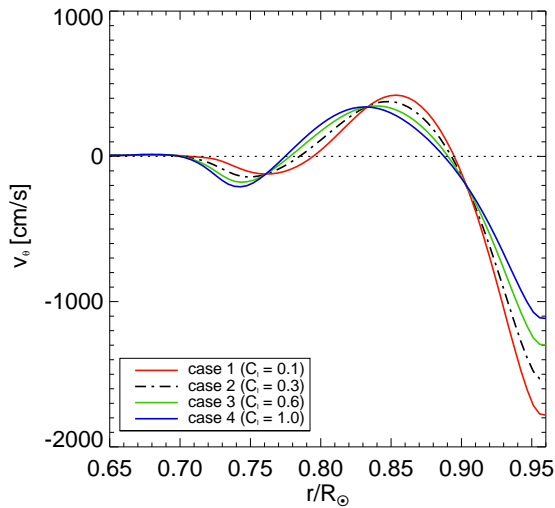}
\caption{Profiles of latitudinal velocity $v_{\theta}$ as functions of radial distance averaged over the middle latitudinal bands $35^{\circ}\mathrm{N} - 45^{\circ}\mathrm{N}$. Black dash-dot line denotes the reference case.} \label{fig:parameter}
\end{figure}

In this section, we discuss the dependence on the free parameter $C_{\mathrm{l}}$ which specifies the magnitude ratio of the $\Lambda$-effect between lower and upper convection zone.
Larger $C_{\mathrm{l}}$ corresponds to the larger effect of the upward turbulent angular momentum transport in the lower part of the convection zone with respect to the effect of the downward transport in the upper layer.
When $C_{\mathrm{l}}$ is changed, the amplitude of the differential rotation is directly affected. Thus, $\Lambda_{0}$ is adjusted such that the equatorial rotation rate remains the same with the helioseismic measurement $\approx 460 \ \mathrm{nHz}$ for all cases as shown in Table \ref{table:2}.

For all cases, both the solar-like differential rotation and the double-cell meridional circulation are achieved in a similar way to the reference case.
Figure \ref{fig:drparam} shows that the general profile of differential rotation is not sensitive to the change in the parameter $C_{\mathrm{l}}$, although the magnitude of the rotation rate near the pole is marginally affected. 
Dependence of the meridional flow is shown in Figure \ref{fig:parameter}.
Even though the amplitude of latitudinal velocity $v_{\theta}$ or the depths of the return flows are dependent on $C_{\mathrm{l}}$ and $\Lambda_{0}$, we can confirm the existence of poleward flow at the surface, equatorward flow in the middle layer, and poleward flow again near the base of the convection zone in all cases.
This suggests that the double-cell structure is a very robust result for the meridional flow as long as the Reynolds stress transports the angular momentum upward (downward) in the lower (upper) half of the convection zone. 
Especially, Figure \ref{fig:parameter} shows that, in our mean-field model, this tendency is almost independent of the relative amplitude of the upward turbulent angular momentum flux in the lower convection zone so that double-cell meridional circulation suggested by \citet{zhao2013} can be realized even when the magnitude of the $\Lambda$-effect is $10$ times weaker in the lower convection zone compared with that in the upper convection zone.

Unlike the general tendency of the double-cell structure of the meridional flow, the Taylor-Proudman balance of the differential rotation turns out to highly depend on the change in $C_{\mathrm{l}}$, since the flow speed near the base of the convection zone which plays a crucial role for the thermal wind balance in suppressing the entropy perturbation is sensitive to $C_{\mathrm{l}}$.
In order to quantitatively discuss the Taylor-Proudman balance of the differential rotation, let us define a new parameter $P_{\mathrm{nTP}}$ which describes the morphology of the differential rotation; to what extent the differential rotation deviates from the Taylor-Proudman state,
\begin{eqnarray}
  P_{\mathrm{nTP}}&=&\frac{1}{R_{\odot}^{2}\Omega_{0}^{2}}\int \frac{\partial \Omega_{1}^{2}}{\partial z} dV \nonumber \\
    &=& \frac{1}{R_{\odot}^{2}\Omega_{0}^{2}}\int \left(\cos{\theta} \frac{\partial}{\partial r} -\frac{\sin{\theta}}{r} \frac{\partial}{\partial\theta}  \right) \Omega_{1}^{2} dV. \label{eq:NTP}
\end{eqnarray}
The fourth column in Table \ref{table:2} shows the absolute values of $P_{\mathrm{nTP}}$ for each case.
$P_{\mathrm{nTP}}$ monotonically decreases with the increase in $C_{\mathrm{l}}$.
In other words, the differential rotation becomes close to the Taylor-Proudman state when the relative amplitude of the turbulent angular momentum flux in the lower region increases.

According to the previous discussion in Section \ref{subsec:TPbalance}, the change in the Taylor-Proudman balance would be attributed to the change in the meridional flow.
In order to evaluate the extent of the double-cell structure quantitatively, we further introduce a new parameter $P_{\mathrm{cw}}$ defined by,
\begin{eqnarray}
P_{\mathrm{cw}}&=&\frac{\int \Psi(r,\theta) dS}{\int |\Psi(r,\theta)| dS}, \label{eq:CW}
\end{eqnarray}
where the integral is taken over the meridional plane and $\Psi(r,\theta)$ denotes the stream-function of the meridional circulation defined by,
\begin{eqnarray}
  \rho_{0}\bm{v}_{m}=\nabla\times(\Psi \bm{e}_{\phi}).
\end{eqnarray}
Here, we set $\Psi =0$ at the boundaries of our numerical domain.
When $P_{\mathrm{cw}}$ is $+(-)1$, the meridional circulation is said to be clockwise (counterclockwise), whereas the double-cell meridional circulation consists of both the clockwise cell and the counterclockwise cell with the same circulation strengths when $P_{\mathrm{cw}}$ is zero.
The values of $P_{\mathrm{cw}}$ for each case are shown in the fifth column of Table \ref{table:2}.

When $C_{\mathrm{l}}$ is sufficiently small (case 1), $P_{\mathrm{cw}}$ takes a negative value, meaning that the meridional flow is dominated by the upper counterclockwise circulation cell, which is the direct consequence of the local angular momentum conservation within the upper half of the convection zone.
On the other hand, when the magnitude of the $\Lambda$-effect in the lower half of the convection zone becomes comparable with respect to the magnitude in the upper layer ($C_{\mathrm{l}}\approx 1$), the influence of the lower clockwise circulation cell driven by the upward turbulent angular momentum transport becomes stronger as manifested by the positive value of $P_{\mathrm{cw}}$ in case 3 and 4.
These results indicate that, if the upward turbulent angular momentum transport works effectively in the lower convection zone, the differential rotation becomes close to the Taylor-Proudman state due to the stronger suppression of the entropy perturbation near the equator by fast clockwise circulation cell in the lower convection zone.

\section{Angular Momentum Balance} \label{sec:refgyro}

\begin{figure*}[]
\figurenum{8}
\plotone{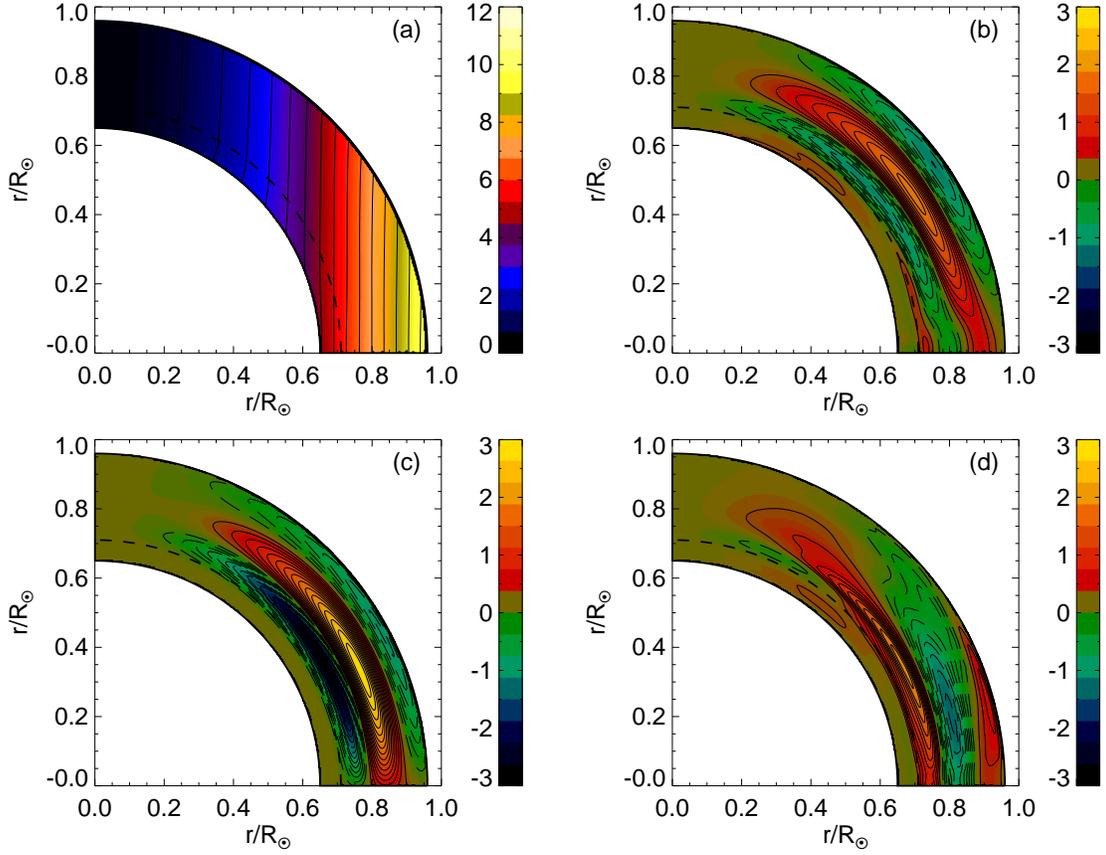}
\caption{Angular momentum balance, emphasizing each term of the equation of gyroscopic pumping (\ref{eq:gyro3}). (a) The mean angular momentum density per unit mass $\mathcal{L}$ in units of $10^{15}\ \mathrm{cm}^{2} \ \mathrm{s}^{-1}$. (b) The angular momentum advected by the meridional flow, which is represented by the left-hand side of the equation (\ref{eq:gyro3}). (c) The turbulent angular momentum transport, which is expressed in the first term on the right-hand side of the equation (\ref{eq:gyro3}). (d) The angular momentum diffusion, which is expressed in the second term on the right hand side of the equation (\ref{eq:gyro3}). Figures (b), (c), and (d) are shown in units of $10^{6}\ \mathrm{g}\ \mathrm{cm}^{-1} \ \mathrm{s}^{2}$.} \label{fig:gyroscopic}
\end{figure*}

In this section, we discuss the angular momentum balance in a dynamical stationary state for the reference solution (case 2), based on the equation of gyroscopic pumping (equation (\ref{eq:gyro1})).
Note that in the stationary state the equation of continuity reduces to the anelastic condition $\nabla\cdot(\rho_{0}\bm{v}_{m})=0$.
The Reynolds stress on the right hand side of the equation (\ref{eq:gyro1}) can be divided into the non-diffusive part ($\Lambda$-effect) and the diffusive part. Therefore, we have,
\begin{eqnarray}
  \rho_{0}\bm{v}_{m}\cdot\nabla\mathcal{L}&=&\nabla\cdot\left[\rho_{0}r\sin{\theta}\nu_{\mathrm{l}}\bm{\Lambda}\Omega_{0}\right] \nonumber \\
  &&+\nabla\cdot\left[\rho_{0}r^{2}\sin^{2}{\theta}\nu_{\mathrm{v}}\nabla\Omega_{1}\right]. \label{eq:gyro3}
\end{eqnarray}
Figure \ref{fig:gyroscopic} illustrates the angular momentum balance in the stationary state, as expressed in each term of the above equation (\ref{eq:gyro3}).
Figure \ref{fig:gyroscopic}(a) shows the mean angular momentum density $\mathcal{L}$ whose contour lines are parallel to the rotational axis, showing that the angular momentum distribution is almost the same with that of solid body rotation.
Figure \ref{fig:gyroscopic}(b) shows the advection of the angular momentum by the meridional circulation (the left-hand-side term of equation (\ref{eq:gyro3})).
Figure \ref{fig:gyroscopic}(c) shows the turbulent angular momentum transport by the non-diffusive part of the Reynolds stress (the first term on the right hand side of equation (\ref{eq:gyro3})), which clearly manifests that the angular momentum is transported from the lower and upper convection zone to the middle layer.
Figure \ref{fig:gyroscopic}(d) shows the dissipative effect of the angular momentum due to the turbulent viscosity (the second term on the right hand side of equation (\ref{eq:gyro3})).

In the stationary state, the advective effect shown in Figure \ref{fig:gyroscopic}(b) is totally balanced by a combination of the $\Lambda$-effect (Figure \ref{fig:gyroscopic}(c)) and the turbulent diffusion (Figure \ref{fig:gyroscopic}(d)).
Figure \ref{fig:gyroscopic}(c) shows that the distribution of the angular momentum is largely changed by the $\Lambda$-effect especially in low to middle latitudes where the momentum arm $\zeta=r\sin{\theta}$ is relatively large.
Figure \ref{fig:gyroscopic}(d), on the other hand, states that the turbulent diffusion also works effectively in this low to middle latitudinal region where the multiplication of $\zeta^{2}$ becomes dominant, and consequently, the tachocline in low latitudes becomes highly subject to the turbulent diffusion.
This means that the turbulent diffusion plays an important role in suppressing the $\Lambda$-effect, leading to a significant amount of cancellation between Figure \ref{fig:gyroscopic}(c) and \ref{fig:gyroscopic}(d).
As a result, we can confirm that there exists a qualitative correspondence between the angular momentum advected by the meridional flow (Figure \ref{fig:gyroscopic}(b)) and the angular momentum transport by the $\Lambda$-effect (Figure \ref{fig:gyroscopic}(c)).
This manifests that the structure of the meridional circulation might be almost determined by the $\Lambda$-effect such that the advection by the meridional flow can balance with the angular momentum transported by the $\Lambda$-effect which is partially suppressed by the turbulent diffusion.
Comparison between Figure \ref{fig:gyroscopic}(b) and \ref{fig:gyroscopic}(c) further shows that the accumulated angular momentum in the middle convection zone due to the $\Lambda$-effect is redistributed to both lower and upper layers of the convection zone by the lower and upper circulation cells, respectively.

\section{Implication for Solar Meridional Circulation} \label{sec:validity}


\begin{figure*}[]
\figurenum{9}
\plotone{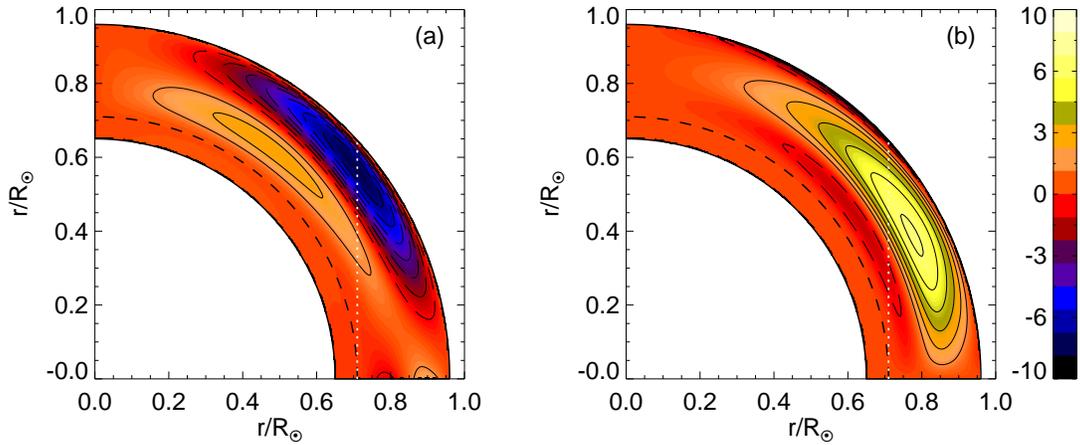}
\caption{Values of (a) $\langle v_{r}'v_{\phi}' \rangle$ and (b) $\langle v_{\theta}'v_{\phi}' \rangle$ obtained in our reference solution. Both values are shown in the meridional plane in units of $10^{6}\ \mathrm{cm^{2}\ s^{-2}}$. White dotted lines indicate the location of the tangential cylinder, $r\sin{\theta}=r_{\mathrm{bc}}$. } \label{fig:correlation}
\end{figure*}


In our mean-field model, the $\Lambda$-effect, which drives the whole large-scale dynamics, is given by a simple consideration of the equation of gyroscopic pumping (equation (\ref{eq:gyro2})) in an ad-hoc manner (Subsection \ref{subsec:lambda-modeling}).
In this section, we briefly discuss the validity of our parameterization of the $\Lambda$-effect and consider to what extent our model is applicable to the solar convection zone.
Here, we should note that our parameterization of the $\Lambda$-effect might not be an unique solution for the double-cell type meridional circulation.\deleted{The change in the $\Lambda$-effect, $\bm{\Lambda} \to \bm{\Lambda}+\tilde{\bm{\Lambda}}$, does not modify the way of the turbulent angular momentum transport as long as the divergence-free condition, $\nabla\cdot(\rho_{0}r\sin{\theta}\nu_{\mathrm{l}}\tilde{\bm{\Lambda}}\Omega_{0})=0$, is satisfied.}\added{However, mainly due to the reasons described in Section \ref{subsec:lambda-modeling},}
we do not\deleted{, therefore,} aim to discuss the general properties of all the possible $\Lambda$-effect in this section but rather focus only on the parameterization of our reference case as an example, from which we can still derive some basic suggestions on the solar meridional circulation structure.

Figure \ref{fig:correlation} shows the values of $\langle v_{r}'v_{\phi}' \rangle$ and $\langle v_{\theta}'v_{\phi}' \rangle$ in the stationary state for the reference solution (case 2) calculated by,
\begin{eqnarray}
  \langle \bm{v}_{m}' v_{\phi}' \rangle =-\left( \nu_{\mathrm{v}}r\sin{\theta}\nabla\Omega_{1} + \nu_{\mathrm{l}}\bm{\Lambda}\Omega_{0} \right),
\end{eqnarray}
which largely reflect our parameterization of the $\Lambda$-effect.
Positive (negative) $\langle v_{r}'v_{\phi}' \rangle$ in the lower (upper) convection zone expresses that the Reynolds stress transports the angular momentum upward (downward) in the lower (upper) layer whereas the positive $\langle v_{\theta}'v_{\phi}' \rangle$ in the bulk of the convection zone describes the equatorward angular momentum transport.
It is concluded, from the discussion in Section \ref{sec:two}, that the positive (negative) $\langle v_{r}'v_{\phi}' \rangle$ in the lower (upper) convection zone is crucial for driving the double-cell type meridional circulation.
Dominant positive $\langle v_{\theta}'v_{\phi}' \rangle$, on the other hand, is needed to accelerate the equatorial region and to obtain the solar-like differential rotation.
In this section, we consider that $\langle v_{r}'v_{\phi}' \rangle$ and $\langle v_{\theta}'v_{\phi}' \rangle$ represent the correlations of the small-scale turbulent flows which cannot be resolved in this mean-field model.
The aim of the following discussion is, therefore, to present possible physical processes which can produce the turbulent correlations described above in a qualitative way and consider if these physical processes exist in the solar convection zone.

Let us first consider the positive $\langle v_{\theta}'v_{\phi}' \rangle$ which is mostly confined outside the tangential cylinder (Figure \ref{fig:correlation}(b)).
It is widely believed that convective columns, known as ``banana cells'' that are mostly aligned with the rotational axis extending across the equator, dominate the convective structure in this region under the strong influence of rotation \citep[e.g.,][]{Busse1970,miesch2005review}.
These banana cells have been repeatedly revealed by 3D full spherical simulations as a coherent north-south alignment of penetrative downflow lanes \citep{brun2004,miesch2006,miesch2008,kapyla2011,featherstone2015,hotta2015}.
As a result, there appears a coherent azimuthal velocity field towards these downflow lanes in the outer layer of the banana cells.
It is expected that the positive correlation $\langle v_{\theta}'v_{\phi}' \rangle$ is established by the Coriolis forces acting on the azimuthal velocity directing the downflow lanes:
The prograde (retrograde) azimuthal flow $v_{\phi}'>0 \ (<0)$ is bent equatorward (poleward) to obtain a positive (negative) latitudinal component, leading to $\langle v_{\theta}'v_{\phi}' \rangle >0$.
\added{This positive $\langle v_{\theta}'v_{\phi}' \rangle$ in the northern hemisphere is confirmed not only by many global simulations listed above but also by observations with different methods; motion tracing of sunspots or sunspot groups \citep{ward1965,gilman1984,pulkkinen1998,sudar2014} and supergranulation tracking \citep{hathaway2013}. In our model, the correlation has its peak at about $30^{\circ}$ similarly to \citet{sudar2014}'s measurements and the profiles obtained by direct or mean-field simulations \citep{kapyla2011,rudiger2014}.}

Next, negative $\langle v_{r}'v_{\phi}' \rangle$ in the upper convection zone ($r>r_{\mathrm{mid}}$) is considered (Figure \ref{fig:correlation}(a)).\deleted{Since convective velocity $v'$ and pressure scale height $H_{p}$ becomes larger and shorter respectively towards the surface according to 3D simulations, the Rossby number $Ro=v'/(2\Omega_{0}H_{p})$ is expected to become relatively large in this region.}In this upper region, \added{since $Ro$ is expected to become relatively large,} the effect of the Coriolis force becomes small and the convectional structure is mostly dominated by the radial motions of the thermal convection.
Negative $\langle v_{r}'v_{\phi}' \rangle$ can be explained by the Coriolis forces acting on these radial flows: The Coriolis force deflects downflows ($v_{r}'<0$) in a prograde way ($v_{\phi}'>0$), inducing $\langle v_{r}'v_{\phi}' \rangle < 0$.
In fact, the inward angular momentum transport driven by this negative $\langle v_{r}'v_{\phi}' \rangle$ is considered as one of the main formation mechanisms of the solar near surface shear layer \citep{foukal1975,miesch2011}, which exists above $r \approx 0.96 \ R_{\odot}$ \citep{howe2011}.
In order to evaluate to what extent the generation process of this negative $\langle v_{r}'v_{\phi}' \rangle$ can prevail in the deeper convection zone, it is necessary to investigate\deleted{the radial dependence of $Ro$; especially} the depth \replaced{above which}{where} $Ro$ exceeds the unity and the influence of the Coriolis force becomes weak.
In \citet{hotta2015}'s recent high-resolution calculations of the solar global convection in which the near surface shear layer is produced above $r=0.95 \ R_{\odot}$, the Rossby-unity line is located about $r \approx 0.93 \ R_{\odot}$.
\replaced{However,}{Although} the profile of $Ro$ presented in their paper may be different from the solar profile because the luminosity is artificially reduced in order to decrease $Ro$ and to obtain a solar-like differential rotation profile,
\replaced{It is also shown}{it should be noted that} in their simulations the negative $\langle v_{r}'v_{\phi}' \rangle$ is not confined to the near surface shear layer but extends down to the middle convection zone $r \approx 0.85 \ R_{\odot}$.
Therefore, it is likely that the negative $\langle v_{r}'v_{\phi}' \rangle$\added{, whose existence at the surface is indirectly supported by measurements of near surface shear layer and poleward meridional flow,} prevails in the upper half of the convection zone, not being confined only to the top surface, as presented in Figure \ref{fig:correlation}(a).

Finally, positive $\langle v_{r}'v_{\phi}' \rangle$ in the lower convection zone is considered (Figure \ref{fig:correlation}(a)).
We should note at first that the amplitude of the positive $\langle v_{r}'v_{\phi}' \rangle$ in the lower convection zone is about three times smaller than the positive one in the upper convection zone in our reference case, and also that it can be even smaller as in case 1.
Apart from the amplitude, it is confirmed that positive $\langle v_{r}'v_{\phi}' \rangle$ is mainly confined inside the tangential cylinder in middle to high latitudes\added{, contrary to the results obtained by direct numerical simulations where positive $\langle v_{r}'v_{\phi}' \rangle$ mostly exists outside the cylinder \citep{kapyla2011,hotta2014}}.
Although the positive $\langle v_{r}'v_{\phi}' \rangle$ outside the tangential cylinder can be explained by the rotation-induced prograde inclination of the banana-cell structure \citep{busse2002,aurnou2007,gastine2013}, the turbulent correlations inside the tangential cylinder cannot be attributed to the properties of the banana cells.
In fact, it is unclear what kind of physical process can generate the positive $\langle v_{r}'v_{\phi}' \rangle$ in the lower convection zone in high latitudinal region.
Yet, our results might suggest that the realization of the double-cell meridional circulation as suggested by \citet{zhao2013} needs the upward transport of the angular momentum by the Reynolds stress even inside the tangential cylinder where the positive turbulent correlation $\langle v_{r}'v_{\phi}' \rangle$ should be explained by some other physical processes but the rotation-induced banana-cell structure.

\section{Summary} \label{sec:summary}

We have investigated the condition of the Reynolds stress needed for maintaining the double-cell meridional circulation which has recently been revealed by helioseismology \citep{zhao2013,kholikov2014}.
We further examined whether or not the double-cell structure of the meridional circulation can coexist with the solar-like differential rotation in the mean-field framework.
This work is significant because the discovery of the double-cell meridional circulation by \citet{zhao2013} is one of the most controversial issues on the solar flux-transport dynamo model \citep{wang89,choudhuri95,hazra2014} yet no one investigated whether or not it is a possible structure from a hydrodynamical perspective.
Our work can be used as a base for further research on the validity of the conventional flux-transport dynamo model.

First, we consider the angular momentum balance for a stationary large scale flows in the solar convection zone, known as gyroscopic pumping \citep{miesch2011} and apply it to the double-cell type meridional flow to derive the properties of the Reynolds stress necessary for driving the double-cell flow structure in a qualitative way.
As a result, positive (nagative) $\langle v_{r}'v_{\phi}' \rangle$ in the lower (upper) convection zone turns out to be indispensable for the double-cell type meridional flow. In other words, the Reynolds stress must transport the angular momentum radially upward (downward) in the lower (upper) half of the convection zone.

Since it is not so obvious whether the Reynolds stress can drive at the same time both the solar-like differential rotation and the double-cell type meridional circulation that are in a complicated hydrodynamical balance \citep{miesch2005review}, we then conduct a set of mean-field simulations where the $\Lambda$-effect \citep{rudiger93,kichatinovrudiger93} is given explicitly such that it reflects the condition of the Reynolds stress described above.
As a consequence, it is confirmed that the double-cell structure of the meridional circulation is hydrodynamically compatible with the solar-like differential rotation as shown in Figure \ref{fig:dr} and Figure \ref{fig:mc}.
We further conduct a parameter survey on the relative magnitude of the $\Lambda$-effect in the lower part of the convection zone with respect to that in the upper part.
Regardless of the amplitude ratio of the upward and downward turbulent angular momentum flux, the double-cell meridional circulation is obtained in all cases along with the solar-like differential rotation, suggesting that just a small fraction of positive turbulent correlation $\langle v_{r}'v_{\phi}' \rangle$ in the lower convection zone ($\approx 10\%$ with respect to the upper layer) is sufficient for attaining the double-cell meridional flow structure.

It is also found that the Taylor-Proudman balance of the differential rotation is mainly influenced by the meridional flow structure.
In the double-cell case, the profile of the differential rotation becomes close to the Taylor-Proudman state compared with the single-cell case \citep{rempel2005}.
It is finally concluded from the results of our parameter survey that the radially inward meridional flow near the equator of the lower clockwise circulation cell suppresses the entropy perturbation generated at the subadiabatic tachocline from spreading out into the convection zone, and therefore, the effect of the thermal wind driven by the latitudinal entropy gradient decreases in our double-cell case.

Next, we again consider the angular momentum balance in a stationary state by evaluating the effects of each term of the equation of gyroscopic pumping (equation (\ref{eq:gyro3})) using our simulational results.
It is found that the $\Lambda$-effect (the first term of the right hand side of the equation (\ref{eq:gyro3})) is almost balanced by the advective effect of the angular momentum by the meridional flow (the left hand side of the equation (\ref{eq:gyro3})) as shown in Figure \ref{fig:gyroscopic}(b) and \ref{fig:gyroscopic}(c).
Therefore, the structure of the meridional circulation turns out to be mainly determined by the parameterization of the $\Lambda$-effect in our model.
The condition of the Reynolds stress that we derived in Section \ref{sec:two} can relate with the maintenance mechanism of the double-cell meridional circulation as follows:
When the Reynolds stress distributes the angular momentum from the lower and upper parts of the convection zone to the middle layer, the meridional flow is driven such that it advects the accumulated angular momentum in the middle convection zone to the upper and lower layers by two radially distinct circulation cells, respectively.

Finally, the validity of our parameterization of the $\Lambda$-effect, which has some physical uncertainties in our mean-field framework, is addressed.
The negative $\langle v_{r}'v_{\phi}' \rangle$ is likely to be generated by the Coriolis forces acting on radial flows in the upper part of the convection zone where $Ro$ is relatively large.
On the other hand, it is generally expected that the positive $\langle v_{r}'v_{\phi}' \rangle$, which is crucial for driving the clockwise circulation cell in the lower part of the convection zone, is due to the rotation-induced banana cell structure in the low $Ro$ regime.
However, it is found that the positive $\langle v_{r}'v_{\phi}' \rangle$ in the lower convection zone is mostly located inside the tangential cylinder in middle to high latitudes where the banana cells may not exist.
Although it is unclear how this turbulent correlation can be produced, our study suggests that the double-cell type meridional flow needs the upward angular momentum transport even inside the tangential cylinder.
The uncertainties on the small-scale convectional structure or the distribution of the turbulent correlations in this region are expected to be resolved using high-resolution global spherical simulations.

Our future work will focus on the effect of the magnetic field on the large-scale flow structures that is not considered in this paper.
Several investigations have been conducted on the feedback of the Lorentz force on differential rotation and the single-cell meridional flow in the mean-field framework but not for the double-cell meridional flow.
\citet{rempel2006a} found that the meridional flow structure shows a deflection from the magnetized region and also that a meridional flow with an amplitude of less than $200 \ \mathrm{cm} \ \mathrm{s}^{-1}$ cannot transport a magnetic field with a strength of $20 -30 \ \mathrm{kG}$ due to the magnetic tension force.
It is, therefore, expected that our lower clockwise cell whose amplitude is about $200 \ \mathrm{cm} \ \mathrm{s}^{-1}$ might be highly affected by the existence of toroidal magnetic fluxes near the base of the convection zone that can be generated by differential rotation through the $\Omega$-effect.
Moreover, the dynamical behavior of the large-scale magnetic field is also expected to depend upon the meridional flow structure.
Since all of the previous studies on the flux-transport dynamo based on the double-cell meridional circulation were carried out in a purely kinematic regime \citep{pipin2013,hazra2014}, our study on the non-kinematic flux-transport dynamo with the double-cell meridional circulation will contribute to the reconsideration of the conventional flux-transport dynamo model in a more realistic regime.\\

We are grateful to Dr.H.Hotta for helpful discussions \added{and an anonymous referee for useful comments on the manuscript}. This work was supported by \added{Leading Graduate Course for Frontiers of Mathematical Sciences and Physics (FMSP)} and JSPS KAKENHI Grant Number 15H03640. Numerical computations were carried out on PC cluster at Center for Computational Astrophysics (CfCA), National Astronomical Observatory of Japan.

\listofchanges

\end{document}